\documentclass{aa}

\usepackage{graphicx}
\usepackage{txfonts}

\begin{document}

\title{VLBI measurement of the secular aberration drift\thanks{We dedicate this work to the memory of Anne-Marie Gontier, our colleague and personal friend, and a widely recognized specialist of VLBI. She passed away shortly after this paper was submitted.}}

\author{O. Titov\inst{1} \and S. B. Lambert\inst{2} \and A.-M. Gontier\inst{2}}

\offprints{O. Titov, \email{Oleg.Titov@ga.gov.au}}

\institute{Geoscience Australia, PO Box 378, Canberra, 2601, Australia
           \and
           Observatoire de Paris, D\'epartement Syst\`emes de R\'ef\'erence Temps Espace (SYRTE), CNRS/UMR8630, 75014 Paris, France}
\date{}

\abstract
{}
{While analyzing decades of very long baseline interferometry (VLBI) data, we detected the secular aberration drift of the extragalatic radio source proper motions caused by the rotation of the Solar System barycenter around the Galactic center. Our results agree with the predicted estimate to be 4--6~micro arcseconds per year ($\mu$as/yr) towards $\alpha=266^{\circ}$ and $\delta=-29^{\circ}$. In addition, we tried to detect the quadrupole systematics of the velocity field.}
{The analysis method consisted of three steps. First, we analyzed geodetic and astrometric VLBI data to produce radio source coordinate time series. Second, we fitted proper motions of 555 sources with long observational histories over the period 1990--2010 to their respective coordinate time series. Finally, we fitted vector spherical harmonic components of degrees 1 and~2 to the proper motion field.}
{Within the error bars, the magnitude and the direction of the dipole component agree with predictions. The dipole vector has an amplitude of $6.4\pm1.5$~$\mu$as/yr and is directed towards equatorial coordinates $\alpha=263^{\circ}$ and $\delta=-20^{\circ}$. The quadrupole component has not been detected. The primordial gravitational wave density, integrated over a range of frequencies less than 10$^{-9}$~Hz, has a limit of $0.0042\;h^{-2}$ where $h$ is the normalized Hubble constant is $H_0/(100~{\rm km/s})$.}
{}

\titlerunning{}

\maketitle

\section{Introduction}

Secular aberration drift is an apparent alteration in the velocity of distant objects caused by the acceleration of the Solar System barycenter in the Milky Way. The proper motion field of extragalactic sources affected by the secular aberration drift should present a dipolar structure with amplitude 4--6~microseconds of arc per year ($\mu$as/yr) directed towards the Galactic center (see, e.g., Fanselow 1983, Bastian~1995, Eubanks et al.~1995, Sovers et al.~1998, Mignard 2002, Kovalevsky 2003, Kopeikin \& Makarov 2006).

Observations of extragalactic radio sources by very long baseline interferometry (VLBI) have been done since the late 1970s by various agencies, mainly the United States Navy and the National Aeronautic and Space Administration (NASA), and have been coordinated since 1998 by the International VLBI Service for Geodesy and Astrometry (IVS, Schl\"uter \& Behrend 2007). The main purpose of these observations is to monitor the Earth orientation and realize terrestrial and celestial reference frames.

One of the cornerstones of the VLBI technique is the accurate realization of the celestial reference system using distant objects that supposedly have no detectable proper motion (Feissel \& Mignard~1998). The most recent realization of the International Celestial Reference Frame (ICRF2, Fey et al.~2009) provides absolute radio source coordinates for 3,414 sources. For 1,448 ICRF2 sources observed in at least two sessions, the median formal error is 175~$\mu$as. For the most observed radio sources, the inflated position errors are 40~$\mu$as in each coordinate.

This catalog did not include proper motions of sources because the dominant proper motion for any source is the somewhat temporary one associated with internal structure changes that produce apparent motions that are an order of magnitude greater than the aberrational drift (e.g., Fey et al. 1997). Such proper motions are not constant with time and are about an order of magnitude higher than the 4--6~$\mu$as/yr proper motions expected from the secular aberration drift. However, we believe that the secular aberration drift can be detected even with source position errors of $\sim$100~$\mu$as and apparent motion formal errors of 10--100~$\mu$as/yr (e.g., Feissel-Vernier 2003).

VLBI measurement of the secular aberration drift is important for both astrometry and astrophysics. Indeed, it would give an independent estimate of the acceleration vector applied to the Solar System barycenter without refering to objects within the Galaxy, so it can be used to constrain the mass of the central regions of the Milky Way. Moreover, the secular aberration drift produces a slow deformation of the celestial reference frame axes, and, if not corrected in geodetic VLBI sofware packages, could lead to contamination in estimates of other geodetic parameters, such as source coordinates and the Earth's nutation and precession (Titov~2010).

Gwinn et al. (1997) have investigated the radio source velocity field obtained from the analysis of 16 years of VLBI data. Though the prime interest of their paper was to find an upper limit for the mass-energy of the cosmological gravitational wave background from the quadrupolar structure of the velocity field, their estimate of the Galactocentric acceleration is higher than expected by a factor of two with comparable standard errors. MacMillan~(2005) processed another seven~years of data but did not detect this effect. Using the OCCAM geodetic VLBI analysis software package, Titov~(2009) reports a statistically significant dipole harmonic, but its magnitude was far from the theoretical value.

The approach used by MacMillan (2005) and Titov (2009) consisted of a direct estimation of the Galactocentric acceleration without intermediate estimation of the proper motions. Our approach differs from these two. We used a three-step procedure that includes (i) a production of source coordinate time series from analysis of VLBI delays, (ii) a least-square fit of proper motions to source coordinate time series, including editing large anomalous proper motions, and (iii) the fit of the Galactocentric acceleration to proper motions. This three-step analysis helped to remove outliers that significantly biased the results of the previous analyses.

\begin{figure}[htbp]
\begin{center}
\includegraphics[width=8.5cm]{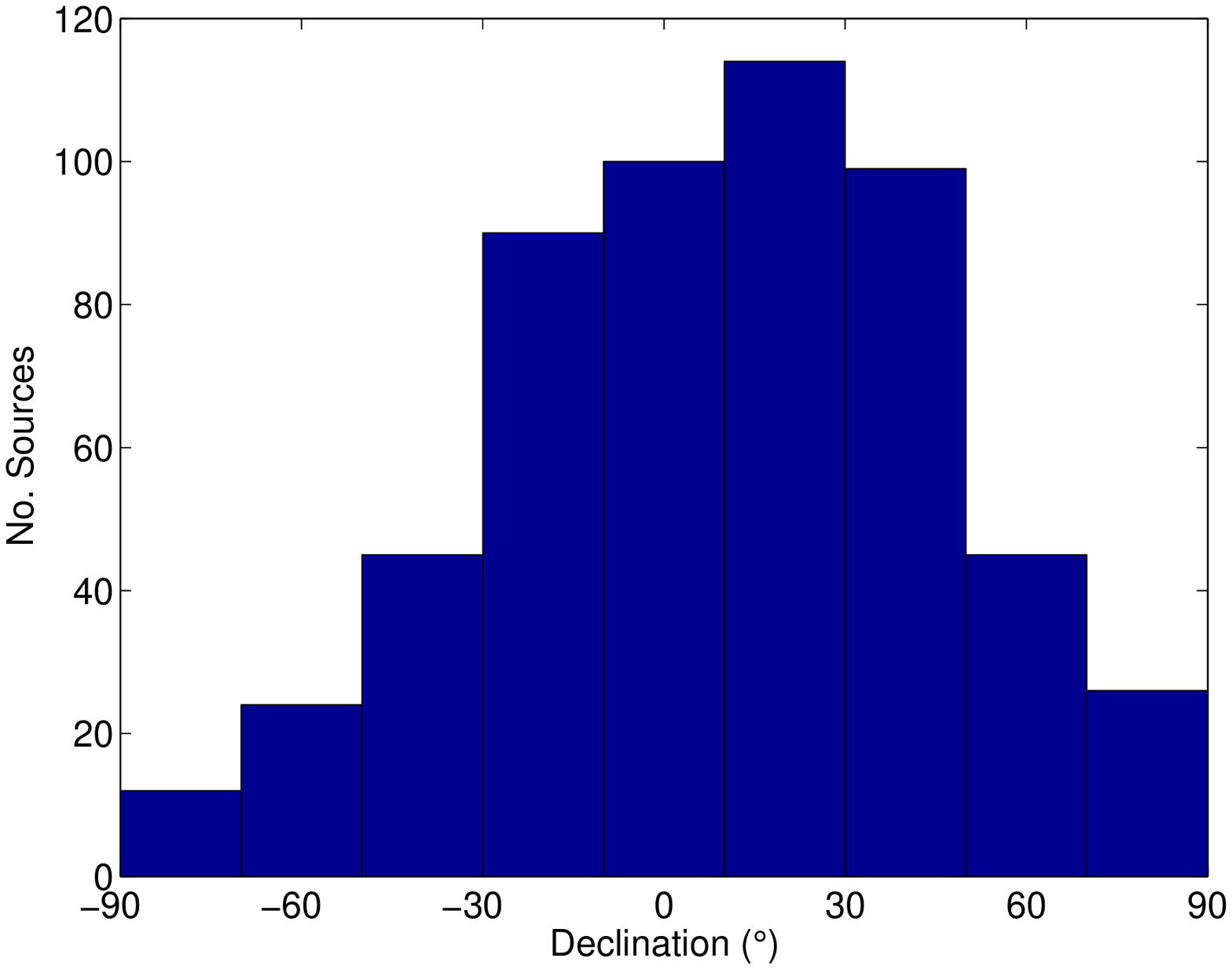}
\includegraphics[width=8.5cm]{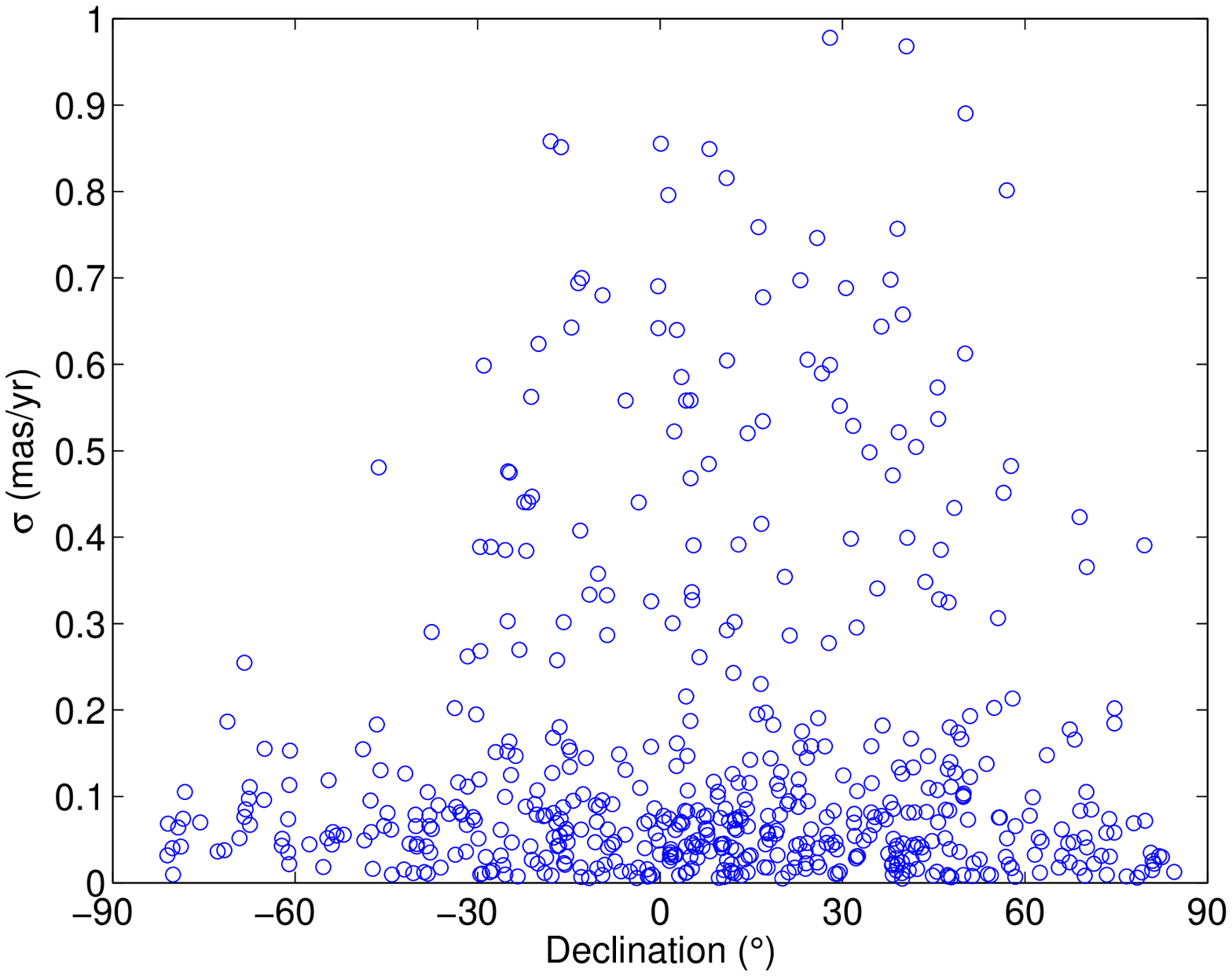}
\end{center}
\caption{The distribution of sources vs. the declination ({\it Top}), the proper motion formal errors $\sigma=(\sigma_{\alpha\cos\delta}^2+\sigma_{\delta}^2)^{1/2}$ vs. the declination ({\it Bottom}).}
\label{fig00}
\end{figure}

\begin{figure}[htbp]
\begin{center}
\includegraphics[width=8.5cm]{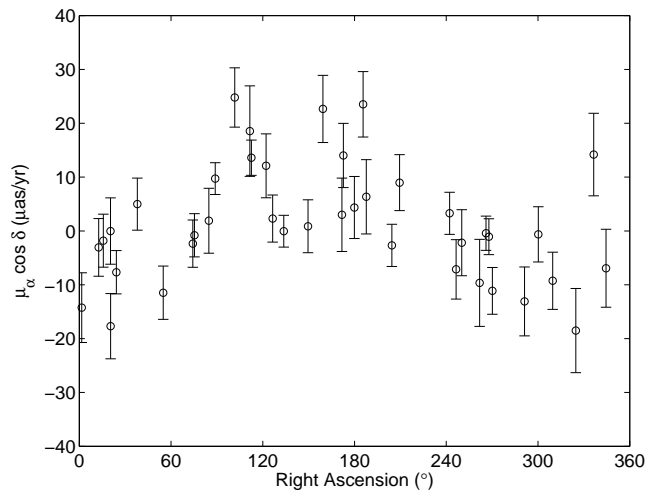}
\end{center}
\caption{The proper motion in right ascension vs. the right ascension of 40 sources observed in more than 1,000 sessions.}
\label{fig20}
\end{figure}

\section{Aberration in proper motions}

Consider the Solar System barycenter moving on a quasi circular orbit at distance $\vec R$ of the Galactic center with an acceleration
\begin{equation}
\vec a=\frac{{\rm d}^2\vec R}{{\rm d}t^2}=-\frac{kM(R)}{R^3}\vec R,
\end{equation}
where $k$ is the constant of gravitation, and $M(R)$ the equivalent mass entering the problem. Using $kM=\omega^2R^3=V^2R$, one gets the acceleration $\vec a={\rm d}\vec V/{\rm dt}=-\omega V\vec u$, resulting in the aberration effect on the proper motion $\vec\mu$ of distant bodies as seen from the Solar System barycenter (e.g., Kovalevsky 2003)
\begin{equation} \label{abb}
\Delta\vec\mu=\frac{\omega V}{c}\vec u,
\end{equation}
wherein $c$ is the speed of light and $\vec u$ the unit vector pointing towards the Galactic center.

Recent estimates of the Galactic parameters give a distance to the Galactic center of $8.4\pm0.6$~kpc and a circular rotation speed of $254\pm16$~km/s (Reid et al.~2009). The expected acceleration $a=V^2/R$ equals $(2.5\pm0.5)\times10^{-13}$~km/s$^2$. This corresponds to an aberration of a distant body proper motion up to $5.0\pm1.0$~$\mu$as/yr.

Further accelerated motion of the whole Milky Way would produce an additional aberration. According to Doppler shift measurements of the Cosmic Background Explorer (COBE) satellite, which realizes a celestial reference frame based on the cosmic microwave background (CMB) radiation, the Local Group of galaxies moves at $\sim$630~km/s (Kogut et al.~1993). Although considered as linear, this motion should correspond to a slowly accelerated motion on longer time scales, of orbit radius of several Mpc. It would result in an aberration that is far too small to be investigated here.

For a distant body of equatorial coordinates $(\alpha,\delta)$, (\ref{abb}) also reads as
\begin{eqnarray} \label{dip}
\Delta\mu_{\alpha}\cos\delta&=&-d_1\sin\alpha+d_2\cos\alpha, \\
\Delta\mu_{\delta}&=&-d_1\cos\alpha\sin\delta-d_2\sin\alpha\sin\delta+d_3\cos\delta,
\end{eqnarray}
where the $d_i$ are the components of the acceleration vector in unit of the proper motion, and which corresponds to degree 1 spheroidal (or electric) development of (see, e.g., Mathews~1981, Mignard \& Morando 1990)
\begin{equation} \label{gen}
\vec\mu=\sum_{l,m}\left(a_{l,m}^E\vec Y_{l,m}^E+a_{l,m}^M\vec Y_{l,m}^M\right),
\end{equation}
where $d_1=a_{1,1}^E$, $d_2=a_{1,-1}^E$, $d_3=a_{1,0}^E$, and $Y_{l,m}^E$ and $Y_{l,m}^M$ are the vector spherical harmonics of electric and magnetic types of degree $l$ and order~$m$.

In addition to the aberration distortion, there may also be a small global rotation that can be described by the toroidal (or magnetic) harmonics of degree~1:
\begin{eqnarray} \label{rot}
\Delta\mu_{\alpha}\cos\delta&=&r_1\cos\alpha\sin\delta+r_2\sin\alpha\sin\delta-r_3\cos\delta, \\
\Delta\mu_{\delta}&=&-r_1\sin\alpha+r_2\cos\alpha,
\end{eqnarray}
where $r_i$ can be expressed in terms of vector spherical harmonics coefficients as $r_1=a_{1,1}^M$, $r_2=a_{1,-1}^M$, and $r_3=a_{1,0}^M$.

To investigate a possible quadrupolar anisotropy of the velocity field, we give the development of the degree 2 vector spherical harmonics (i.e., $l=2$ in Eq.~(\ref{gen})):
\begin{eqnarray} \label{quad}
\Delta\mu_{\alpha}\cos\delta&=&-(a_{2,2}^{E,\rm Re}\sin 2\alpha-a_{2,2}^{E,\rm Im}\cos 2\alpha)\cos\delta \nonumber \\
                            & &+(a_{2,1}^{E,\rm Re}\sin\alpha-a_{2,1}^{E,\rm Im}\cos\alpha)\sin\delta \nonumber \\
                            & &+(a_{2,2}^{M,\rm Re}\sin 2\alpha-a_{2,2}^{M,\rm Im}\cos 2\alpha)\sin\delta\cos\delta \nonumber \\
                            & &+(a_{2,1}^{M,\rm Re}\sin\alpha-a_{2,1}^{M,\rm Im}\cos\alpha)\cos2\delta \nonumber \\
                            & &-a_{2,0}^M\sin\delta\cos\delta, \\
\Delta\mu_{\delta}&=&-(a_{2,2}^{E,\rm Re}\cos 2\alpha+a_{2,2}^{E,\rm Im}\sin 2\alpha)\sin\delta\cos\delta \nonumber \\
                  & &-(a_{2,1}^{E,\rm Re}\cos\alpha+a_{2,1}^{E,\rm Im}\sin\alpha)\cos 2\delta \nonumber \\
                  & &+(a_{2,2}^{M,\rm Re}\cos 2\alpha+a_{2,2}^{M,\rm Im}\sin 2\alpha)\cos\delta \nonumber \\
                  & &-(a_{2,1}^{M,\rm Re}\cos\alpha+a_{2,1}^{M,\rm Im}\sin\alpha)\sin\delta \nonumber \\
                  & &+a_{2,0}^E\sin\delta\cos\delta.
\end{eqnarray}

\section{Results and discussion}

\subsection{Data processing}

We processed 5,030 sessions of the permanent geodetic and astrometric VLBI program since 1979, totalling 7,285,312 group delay measurements at 8.4~GHz. Radio source coordinates were estimated once per session, together with Earth orientation parameters and station coordinates. The cut-off elevation angle was set to 5$^{\circ}$. A priori zenith delays were determined from local pressure values (Saastamoinen 1972), which were then mapped to the elevation of the observation using the Vienna mapping functions (B\"ohm et al.~2006). Zenith wet delays were estimated as a continuous piecewise linear function at 30-min intervals. Troposphere gradients were estimated as 8-hr east and north piecewise functions at all stations except a set of 110 stations with short observational histories. Station heights were corrected for atmospheric pressure and oceanic tidal loading. The relevant loading quantities were deduced from surface pressure grids from the U.~S. NCEP/NCAR reanalysis project atmospheric global circulation model (Kalnay et al.~1996, Petrov \& Boy~2004) and from the FES~2004 ocean tide model (Lyard et al.~2004). No-net rotation (NNR) and translation constraints per session were applied to the positions of all stations, excluding Fort Davis (Texas), Pie Town (New Mexico), Fairbanks (Alaska), and the TIGO antenna at Concepci\'on, Chile because of strong non linear displacements (These two sites experienced post-seismic relaxation effects after large earthquakes on the Denali fault in 2003, and between Talca and Concepci\'on in early 2010). A priori precession and nutation comply with the IAU~2000/2006 resolutions, which include the nutation model of Mathews et al.~(2002), the improved precession model of Capitaine et al.~(2003b), and the non rotating origin-based coordinate transformation between terrestrial and celestial coordinate systems (Capitaine et al.~2003a). Usually, an NNR constraint is used to fix the ICRS axes. However, Titov (2010) argues that application of a tight NNR constraint may wipe out all systemtic effects in the proper motion of reference radio sources. Therefore, we tied the celestial frame to the ICRF2 using a loose NNR constraint uniformly applied for each session. More details are discussed later in Section~3.2. The calculations used the Calc~10.0/Solve~2010.05.21 geodetic VLBI analysis software package, which was developed and maintained at NASA Goddard Space Flight Center, and were carried out at the Paris Observatory IVS Analysis Center (Gontier et al.~2008).

\begin{table*}
\begin{center}
\begin{tabular}{lrrrrr}
\hline
\hline
\noalign{\smallskip}
& DR & DR1 & DR2 & DR3 & DR4 \\
\noalign{\smallskip}
\hline
\noalign{\smallskip}
No. sources & 555 & 268 & 257 & 40 & 515 \\
\noalign{\smallskip}
\hline
\noalign{\smallskip}
Dipole & & & \\
\noalign{\smallskip}
$d_1$      & $-0.7 \pm 0.8$ & $-1.9 \pm 1.0$ & $-1.6 \pm 0.9$ & $-3.4 \pm 1.0$ & $ 4.2 \pm 1.3$ \\
\noalign{\smallskip}
$d_2$      & $-5.9 \pm 0.9$ & $-7.1 \pm 1.1$ & $-6.0 \pm 1.0$ & $-6.2 \pm 1.2$ & $-4.6 \pm 1.3$ \\
\noalign{\smallskip}
$d_3$      & $-2.2 \pm 1.0$ & $-3.5 \pm 1.2$ & $-2.9 \pm 1.1$ & $-3.8 \pm 1.2$ & $ 1.4 \pm 1.7$ \\
\noalign{\smallskip}
Amplitude  & $ 6.4 \pm 1.5$ & $ 8.1 \pm 1.9$ & $ 6.9 \pm 1.7$ & $ 8.0 \pm 2.0$ & $ 6.4 \pm 2.5$ \\
\noalign{\smallskip}
\hline
\noalign{\smallskip}
Rotation & & & \\
\noalign{\smallskip}
$r_1$      & $-2.6 \pm 0.9$ & $-3.2 \pm 1.2$ & $-2.7 \pm 1.1$ & $-1.9 \pm 1.3$ & $-4.3 \pm 1.4$ \\
\noalign{\smallskip}
$r_2$      & $ 0.4 \pm 1.0$ & $-0.6 \pm 1.2$ & $-0.3 \pm 1.1$ & $ 2.5 \pm 1.4$ & $-2.1 \pm 1.5$ \\
\noalign{\smallskip}
$r_3$      & $ 0.8 \pm 0.7$ & $-0.3 \pm 0.9$ & $ 0.0 \pm 0.8$ & $-0.5 \pm 0.9$ & $ 4.1 \pm 1.2$ \\
\noalign{\smallskip}
Amplitude  & $ 2.8 \pm 1.5$ & $ 3.3 \pm 1.9$ & $ 2.7 \pm 1.7$ & $ 3.2 \pm 2.0$ & $ 6.3 \pm 2.4$ \\
\noalign{\smallskip}
\hline
\noalign{\smallskip}
\multicolumn{2}{l}{Direction of the acceleration vector} & & & & \\
\noalign{\smallskip}
Right Ascension ($^\circ$) & $263\pm11$ & $255\pm11$ & $255\pm11$ & $241\pm12$ & $312\pm16$ \\
\noalign{\smallskip}
Declination ($^\circ$) & $-20\pm12$ & $-25\pm11$ & $-25\pm11$ & $-28\pm12$ & $13\pm21$ \\
\noalign{\smallskip}
\hline
\noalign{\smallskip}
Pre-fit wrms & 22.1 & 19.7 & 16.3 & 9.2 & 33.3 \\
\noalign{\smallskip}
Post-fit wrms & 21.7 & 18.9 & 15.7 & 7.5 & 32.7 \\
\noalign{\smallskip}
Reduced $\chi^2$ & 1.9 & 1.9 & 1.8 & 2.1 & 1.9 \\
\noalign{\smallskip}
\hline
\end{tabular}
\end{center}
\caption{The multipole coefficients ($\mu$as/yr) of the velocity field. Uncertainties are 1$\sigma$.}
\label{tab01}
\end{table*}

Before 1990, the general deficiency of the VLBI networks, including the number of observed sources and observing antennas per session, makes the VLBI products less reliable (see, e.g., Gontier et al. 2001, Malkin 2004, Feissel-Vernier et al. 2004, Lambert \& Gontier 2009 who reports interesting statistical results and remarks about the VLBI evolution over the past two decades). For this reason we removed data before 1990. A treatment of the full data base over 1979--2010 is nevertheless presented later for comparison.

In the coordinate time series, data points resulting from fewer than three reliable observations within a session were removed, and outliers were eliminated so that the $\chi^2$ is reasonably close to unity. Then, proper motions were computed by weighted least-squares for time series containing at least ten points and longer than ten~years. Weights were taken as the inverse of the squared formal error. A set of 39 sources showing significant non linear positional variations due to large-scale variations in their structure  (including 3C84, 3C273B, 3C279, 3C345, 3C454.3, and 4C39.25) were isolated in the ICRF2 work and treated in such a manner that they did not perturb the geodetic solutions (Fey et al.~2009). We removed these 39 sources from our data set.

The final sample contains proper motions of 555 sources and is made available electronically. Figure~\ref{fig00} displays the distribution of sources and proper motion formal errors in declination. Near the polar areas, the number of sources decreases proportionally to the cosine of the declination. It also shows the nonuniformity of the sample and a lack of sources at declinations under $-40^{\circ}$. Figure~\ref{fig20} displays $\mu_{\alpha}\cos\delta$ versus $\alpha$ for 40 sources observed in more than 1,000 sessions (see Section 3.2 for details). The apparent motions of these sources are estimated very accurately thanks to a large number of observations. A systematic in $\sin\alpha$ of magnitude less than 10~$\mu$as clearly shows up. Tiny underlying aberrational drift is indicative even for a limited number of well-observed radio sources.

\subsection{Dipole}

This section comprises our results of the dipole component estimation. We start with a main solution including all 555 radio sources. Then we consider different subsets of radio sources to verify the robustness of the main solution.

\begin{figure}
\begin{center}
\includegraphics[width=8.5cm]{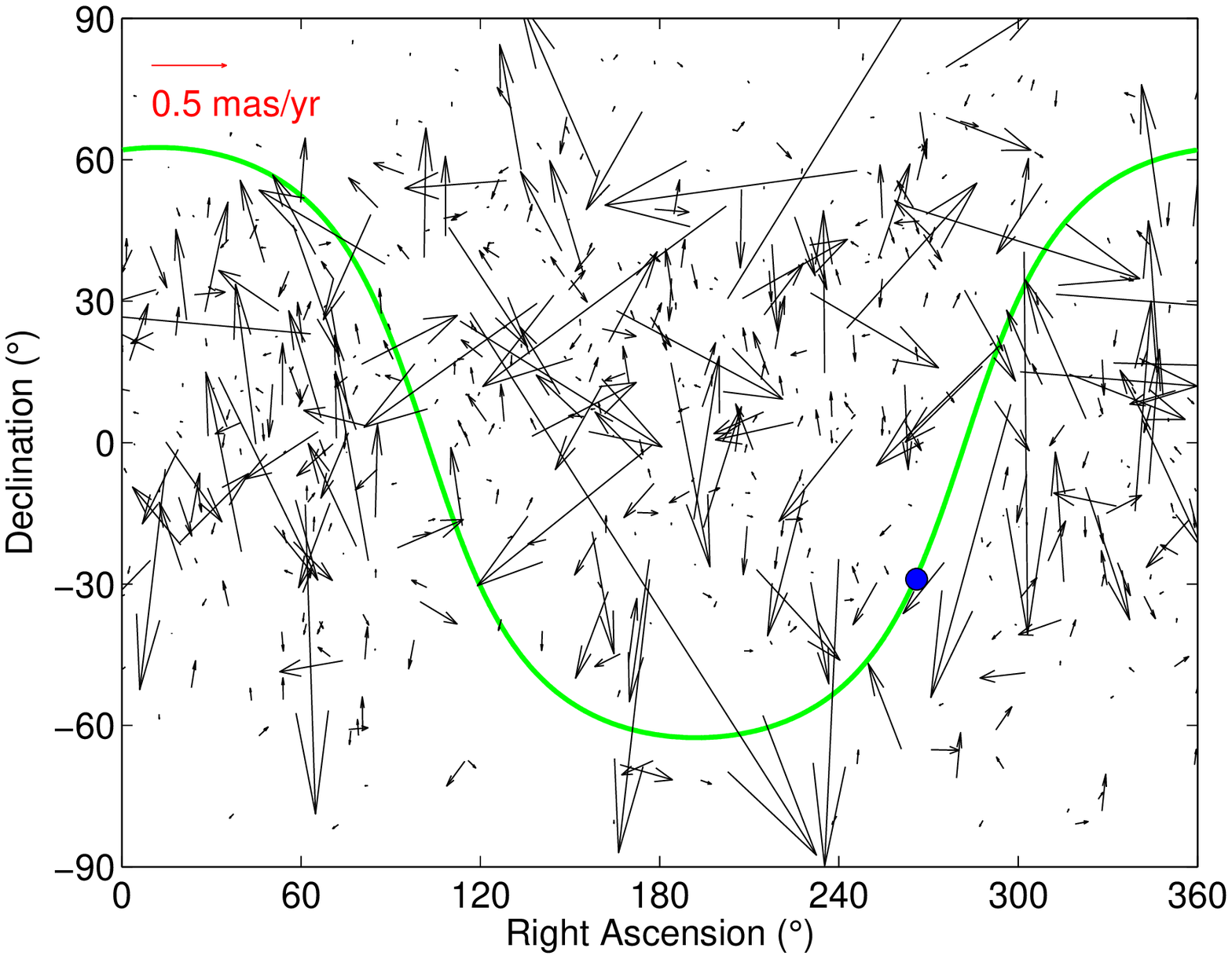}
\includegraphics[width=8.5cm]{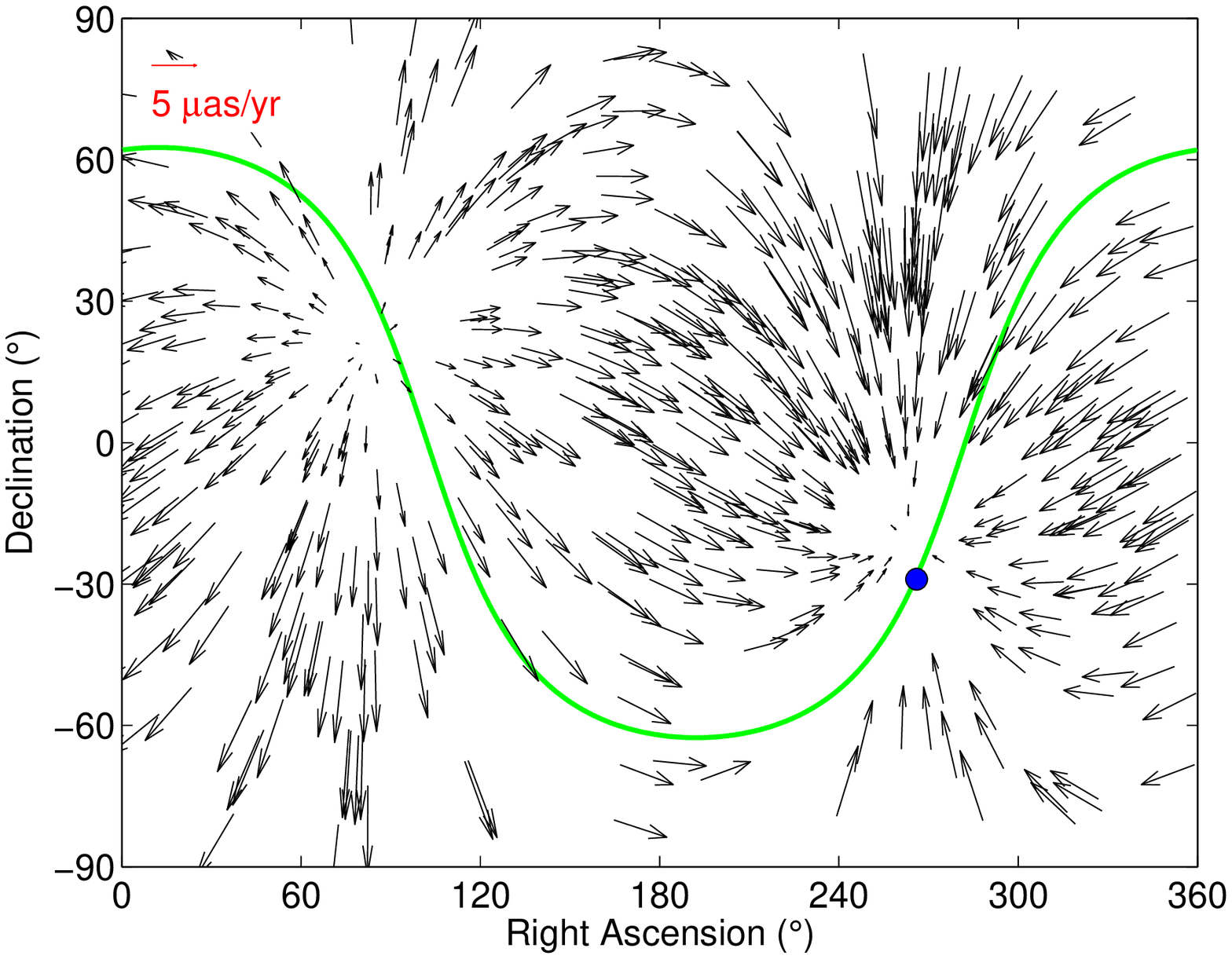}
\end{center}
\caption{({\it Top}) The proper motions of the 555 sources, and ({\it Bottom}) the estimated dipole component of the velocity field. The green line represents the equator of the Milky Way, whose center is indicated by the blue marker.}
\label{fig01}
\end{figure}

First, dipole and rotation coefficients were fitted by weighted least-squares following Eqs.~(3)--(4) and (6)--(7) (Table~\ref{tab01}, column DR). Reported errors are standard formal errors. The fit produces correlations of $\sim$0.4 between $d_1$ and $r_2$ and between $d_2$ and $r_1$. Figure~\ref{fig01} displays both the proper motions of the 555 sources and the estimated dipole component of the velocity field. Within error bars, the dipole amplitude and direction agree with predictions from measurements of the Galactic parameters. The corresponding centrifugal acceleration of the Solar System barycenter is $(3.2\pm0.7)\times10^{-13}$~km/s$^2$. By fixing the distance of the Solar System to the Galactic center to 8.4~kpc, the DR solution yields a circular rotation speed of~276~km/s.

To test the robustness of our estimates, we fitted the parameters to various subsamples of sources, including sources with structure indices (Charlot~1990, Fey \& Charlot 2000) lower than 2.5 (DR1), thus keeping only the most compact sources only. Since structure indices were not available for all sources, the number of sources in the sample is considerably reduced. Nevertheless, the amplitude and the orientation of the dipole do not change significantly with respect to DR. Another test consisted of only keeping ICRF2 defining sources (DR2). Two hundred fifty seven of them have a sufficient observational history to pass our selection scheme successfully. Again, it does not change the amplitude and the orientation of the dipole significantly.

In all the above solutions, the rotation is about 5~$\mu$as/yr and is consistent with the ICRF2 error. If the rotation is fixed to zero, the estimated dipole is of $7.1\pm1.5$~$\mu$as/yr toward ($\alpha=263\pm9^{\circ}$, $\delta=-17\pm11^{\circ}$).

Owing to the uneven number of observations for the 555 radio sources, their proper motion standard errors vary by three orders of magnitude: from 5 to 5000~$\mu$as/yr. It is necessary to check whether the estimates of the dipole component are dominated by a small number of radio sources with small errors on proper motion. Another fit was therefore done to the 40 sources observed in more than 1,000 sessions and used to produce Fig.~\ref{fig20}~(DR3). These sources have formal errors of proper motions less than 15~$\mu$as. DR3 has the best weighted rms among all the solutions in Table~\ref{tab01}.

The last fit (DR4) was done to the remaining 515 sources observed in less than 1,000 sessions. DR3 and DR4 deviate from the main solution DR in opposite directions. Though the orientation of the dipole is mainly constrained by the 40 sources with the longest observational history, the magnitude of the dipole is closer to the theoretical values when the 515 sources are used.

We also made the adjustment without removing data before 1990. It yielded a dipole amplitude of $6.4\pm1.5$~$\mu$as/yr, directed towards ($\alpha=262\pm10^{\circ}$, $\delta=-15\pm12^{\circ}$). Though the amplitude is comparable to the one obtained over 1990--2010, the direction appeared changed by almost 5$^{\circ}$ in declination. It is likely due to a lack of sources observed in the southern hemisphere during the first decade of VLBI.

Overall, the estimate of the dipole effect obtained from the DR solution ($6.4\pm1.5$~$\mu$as/yr; $\alpha=263\pm11^{\circ}$, $\delta=20\pm12^{\circ}$) is considered confident and could be recommended for futher applications.

\subsection{NNR constraint}

Before going further, we address the problem of the constraint that was applied to the observation during the first step of the VLBI data reduction (i.e., production of source coordinate time series). Table~\ref{tab10} displays the amplitude and the orientation of the dipole fitted to time series obtained with different constraints. The first line is the DR solution for which the $\sigma$ of the NNR constraint is 2~as. Decreasing the $\sigma$ to a value close to the milli arcsecond or lower makes the dipole move off the Galactic center.

The last two lines of Table~\ref{tab10} mention solutions without NNR. Applying no constraint at all would cause degeneracy of the normal equation matrix for daily parameters. Therefore, in this solution, each source is tied to the ICRF2 position using a loose constraint of 2~as. We can see that such a constraint is equivalent to the NNR applied with the same $\sigma$.

\begin{table}
\begin{center}
\begin{tabular}{lrrrrr}
\hline
\hline
\noalign{\smallskip}
NNR & $\sigma$ & Amplitude & $\alpha$ & $\delta$ \\
\noalign{\smallskip}
\hline
\noalign{\smallskip}
Yes & 2~as & $6.4\pm1.5$ & $263\pm11$ & $-20\pm12$ \\
\noalign{\smallskip}
Yes & 2~mas & $5.2\pm0.7$ & $237\pm7$ & $22\pm8$ \\
\noalign{\smallskip}
No & 2~as & $6.5\pm1.5$ & $263\pm11$ & $-23\pm12$ \\
\noalign{\smallskip}
No & 2~mas & $3.3\pm0.5$ & $242\pm7$ & $22\pm7$ \\
\noalign{\smallskip}
\hline
\end{tabular}
\end{center}
\caption{The dipole amplitude ($\mu$as/yr) and orientation ($^{\circ}$) obtained with different constraints with uncertainties of 1$\sigma$.}
\label{tab10}
\end{table}

\subsection{Quadrupole}

In a second fit, we adjusted all parameters of Eqs.~(3)--(4) and (6)--(9) (Table~\ref{tab02}). Maximum correlations of 0.5 showed up between $d_1$ and $a_{2,1}^{M,\rm Re}$. Including of the quadrupole harmonics did not change the dipole and rotational effects. The amplitude of the aberrational effect increased by 0.7~$\mu$as/yr, and the right ascension of the vector moved off the theoretical value, but the declination of the vector direction almost coincided now with the theoretical prediction. Fixing the dipole and rotation parameters to the DR values and adjusting the quadrupole parameters only leads to $6.4\pm3.4$~$\mu$as.

The quadrupole component may come either from the Hubble constant anisotropy or the primordial gravitational waves (Kristian \& Sachs~1966, Pyne et al.~1996, Gwinn et al.~1997, Jaffe 2004). Though the Hubble constant anisotropy only affects $a_{2,0}^E$ and $a_{2,2}^{E,\rm Re}$ (Titov~2009), the primordial gravitational waves affect all the quadrupole terms in Table~\ref{tab02}. Whereas the resultant amplitude is less than 3$\sigma$ standard error, the only marginally statistically significant component $a_{2,1}^{E,\rm Re}=4.1\pm1.3$~$\mu$as/yr can be converted into the energy density of gravitational waves $\Omega_{\rm GW}$. Using Eq.~(11) of Gwinn et al.~(1997) for this component, one gets $\Omega_{\rm GW}=0.0042\pm0.0004\;h^{-2}$, where $h=H_0/(100~{\rm km/s})$ is the normalized Hubble constant. The squared proper motion of quasars at cosmologic distances are sensitive to gravitational waves of long wavelength, even comparable to the scale of the Universe, and proportional to $\Omega_{\rm GW}$ on a wide range of frequencies, from the inverse of the period of observations to Hubble time. Therefore, the value estimated above may indicate the upper limit of the gravitational waves density integrated over a range of frequencies less than 10$^{-9}$~Hz.

\subsection{Dependence on the redshift}

\begin{figure}
\begin{center}
\includegraphics[width=8.5cm]{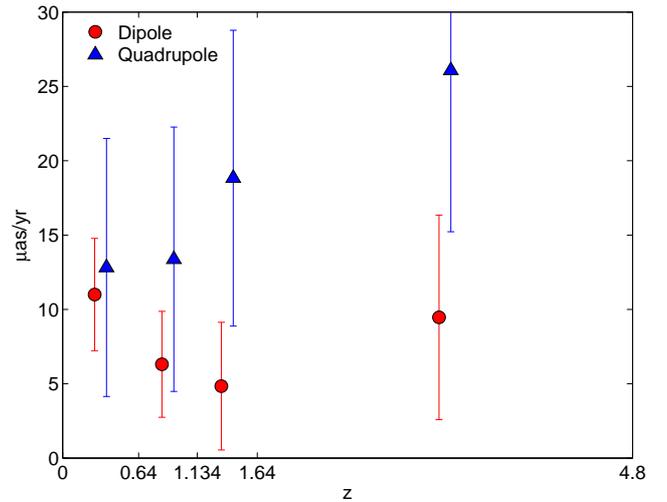}
\end{center}
\caption{The dipole (red disks) and quadrupole (blue triangles) amplitudes as functions of the redshift. Uncertainties are 1$\sigma$.}
\label{fig02}
\end{figure}

We also looked at the dipole and quadrupole amplitudes as functions of the redshift, available for 488 sources (Titov \& Malkin~2009). Each estimate incorporated a subsample containing 122 sources whose redshifts are between the values indicated by the $x$-axis ticks of Fig.~\ref{fig02}. The mean declination of each subsample is between 7.2$^{\circ}$ and 11.7$^{\circ}$. The dipole and quadrupole amplitudes do not present significant dependence on~$z$.

\begin{table}
\begin{center}
\begin{tabular}{lrr}
\hline
\hline
\noalign{\smallskip}
No. sources & 555 & \\
\noalign{\smallskip}
\hline
\noalign{\smallskip}
Dipole & & \\
\noalign{\smallskip}
$d_1$                & $ 0.7 \pm 0.9$ & \\
\noalign{\smallskip}
$d_2$                & $-6.2 \pm 1.0$ & \\
\noalign{\smallskip}
$d_3$                & $-3.3 \pm 1.0$ & \\
\noalign{\smallskip}
Amplitude            & $ 7.1 \pm 1.7$ & \\
\noalign{\smallskip}
\hline
\noalign{\smallskip}
Rotation & & \\
\noalign{\smallskip}
$r_1$                & $-2.4 \pm 1.0$ & \\
\noalign{\smallskip}
$r_2$                & $ 0.4 \pm 1.1$ & \\
\noalign{\smallskip}
$r_3$                & $ 0.8 \pm 0.7$ & \\
\noalign{\smallskip}
Amplitude            & $ 2.6 \pm 1.7$ & \\
\noalign{\smallskip}
\hline
\noalign{\smallskip}
Quadrupole & Re & Im \\
\noalign{\smallskip}
$a_{2,2}^E$ & $ 1.6 \pm 1.0$ & $-0.6 \pm 1.0$ \\
\noalign{\smallskip}
$a_{2,1}^E$ & $ 4.1 \pm 1.3$ & $-1.7 \pm 1.2$ \\
\noalign{\smallskip}
$a_{2,0}^E$ & $ 2.8 \pm 1.2$ & \\
\noalign{\smallskip}
$a_{2,2}^M$ & $-0.6 \pm 1.3$ & $ 2.4 \pm 1.3$ \\
\noalign{\smallskip}
$a_{2,1}^M$ & $ 2.0 \pm 1.1$ & $-0.6 \pm 1.1$ \\
\noalign{\smallskip}
$a_{2,0}^M$ & $ 0.7 \pm 0.9$ & \\
\noalign{\smallskip}
Amplitude   & $ 6.4 \pm 3.6$ & \\
\noalign{\smallskip}
\hline
\noalign{\smallskip}
\multicolumn{2}{l}{Direction of the acceleration vector} & \\
\noalign{\smallskip}
Right Ascension ($^\circ$) & $277\pm12$ & \\
\noalign{\smallskip}
Declination ($^\circ$) & $-28\pm10$ & \\
\noalign{\smallskip}
\hline
\noalign{\smallskip}
Pre-fit wrms & 22.1 & \\
\noalign{\smallskip}
Post-fit wrms & 21.6 & \\
\noalign{\smallskip}
Reduced $\chi^2$ & 1.9 & \\
\noalign{\smallskip}
\hline
\end{tabular}
\end{center}
\caption{The quadrupole coefficients ($\mu$as/yr) of the velocity field with uncertainties od 1$\sigma$.}
\label{tab02}
\end{table}

\section{Conclusion}

This study showed that VLBI has now accumulated accurate enough data to detect the Galactocentric acceleration through its effect on distant radio source positions. It turns out that the current definition of the celestial reference frame as epochless and based on the assumption that quasars have no detectable proper motions should be mitigated. In the future, VLBI realizations of the celestial reference system should correct source coordinates for this effect, possibly by providing source positions, together with a corrective formula.

The European optical astrometry mission Gaia (Perryman et al. 2001), scheduled for 2012, should be able to determine the components of the acceleration vector with a relative precision of 10\% (Mignard 2002). To improve the VLBI determination of the Galactocentric acceleration and to confirm the significance of the quadrupole systematics, more proper motions of extragalactic radio sources need to be measured over the next decade. Concentrating on sources showing a high positional stability and having a low structure index would reduce unwanted effects of intrinsic motion caused by the relativistic jets and other modification of the source structure. In addition, it is necessary to run a dedicated program to measure the redshift of the reference radio sources using large optical facilities, especially in the southern hemisphere (Maslennikov et al.~2010).

\begin{acknowledgements}
The authors are grateful to M. Eubanks, D. Jauncey, S. Klioner, S. Kopeikin, S. Kurdubov, C. Le Poncin-Lafitte, and C. Ma for useful discussions regarding theoretical issues and practical advices on the methods. A special thanks goes to O. Sovers for finding the early report by J. Fanselow mentioning the secular aberration drift. The authors also thank an anonymous referee who helped in improving the manuscript. This paper has been published with permission of the Chief Executive Officer of Geoscience Australia.
\end{acknowledgements}

\end{document}